# Temperature dependence of the EPR linewidth of $Yb^{3+}$ - ions in $Y_{0.99}Yb_{0.01}Ba_2Cu_3O_X$ ($6 \leq X \leq 7$) compounds: Evidence for an anomaly near the superconducting transition

M.R. Gafurov[1,2], L.K. Aminov[1], I.N. Kurkin[1], and V.V. Izotov[1]

[1] *Department of Physics, Kazan State University, 420008 Kazan, Russian Federation*
[2] *I.Physikalisches Institut, Justus-Liebig Universität, D-35392 Giessen, Germany*

## Abstract

Electron paramagnetic resonance experiments on doped $Yb^{3+}$ ions in $Y_{0.99}Yb_{0.01}Ba_2Cu_3O_X$ ($6 \leq X \leq 7$) compounds with different oxygen contents have been made. We have observed the strong temperature dependence of the EPR linewidth in all the investigated samples caused by the Raman processes of spin-lattice relaxation. The spin-lattice relaxation rate anomaly revealed near $T_C$ in the superconducting species can be assigned to the phonon density spectrum changes.

## Introduction

$YBa_2Cu_3O_X$ (YBaCuO, YBCO, 1-2-3) compounds are superconductors at $X > 6.35$ ($T_C = 92$ K for $X = 7.0$) and still intensively studied by the different spectroscopy methods such as inelastic neutron spectroscopy (INS), Mössbauer and Raman spectroscopy, nuclear magnetic (NMR) and electron paramagnetic resonance (EPR) methods [1] in order to accumulate the data for a thorough understanding of the origin of high-temperature superconductivity (HTSC).

The behavior of phonons in high-temperature superconductors is of great significance as dynamic lattice effects may play a relevant role in the mechanism of underlying superconductivity in these compounds. There are only a few publications on electron spin-lattice relaxation (SLR) processes in YBCO (see review [2] and references therein). The most direct and reliable pulse EPR techniques - electron spin-echo and pulsed saturation - are not applicable for this because of very short relaxation times. These measurements, nevertheless, could give information about the spin-lattice relaxation mechanisms, the vibrational spectrum of the substance being studied, and their features.

In this paper, we present the spin-lattice relaxation rate measurements of doped $Yb^{3+}$ ions from the temperature dependence of electron spin resonance linewidth in $Y_{0.99}Yb_{0.01}Ba_2Cu_3O_X$ with different oxygen contents.

**Experimental details**

The polycrystalline powder $YBa_2Cu_3O_X$ samples were prepared by the standard solid-state reaction technique. Appropriate proportions of $Y_2O_3$, $BaCO_3$, $CuO$ were dried at 400 - 500°C, mixed and ground thoroughly into a fine powder. The $Yb^{3+}$ dopants were added using an oxide $Yb_2O_3$ in a ratio Yb:Y = 1:100. These mixtures were converted to $RYBa_2Cu_3O_7$ by thermal treatment, after that the oxygen content was reduced to the specified value X by heating the samples. The exact value of X depends on the annealing procedure and was defined from the lattice parameter along the crystallographic *c*-axis [3] using X-ray diffraction. Purity checking of our samples by means of X-ray phase analysis does not reveal any impurity phases with the accuracy higher than 1%. The values of $T_C$ for different X were determined from the temperature dependence of microwave absorption in a low magnetic field. In present work, we have investigated four samples with different oxygen contents (see Table 1).

It is very important to investigate single crystal samples of YBCO, because of their strong anisotropy. In this work, the YBCO powders were milled (size = 1 – 3 µm), then mixed with paraffin or epoxy resin and placed in a glass tube in a strong magnetic field (≥ 15 kG) to prepare the quasi-single-crystal samples. The *c* – axes of the individual crystallites were predominantly oriented along the direction ***C*** of the aligning magnetic field after hardening of epoxy resin. The EPR linewidth of $Yb^{3+}$ ions in the samples prepared by the procedure described above was narrow enough in the H∥***C*** orientation (see Table 1) to observe the anomaly near $T_C$.

EPR spectra were recorded on X-band (≈ 9.25 GHz) IRES-1003 and THN-251 EPR spectrometers (≈ $10^{-5}$ W) in the temperature range from 10 to 120 K.

**Results and discussion**

Fig. 1 shows the EPR spectrum of $Yb^{3+}$ ions in $Y_{0.99}Yb_{0.01}Ba_2Cu_3O_{6.67}$. In addition, the unavoidable intensive signal at g ≈ 2 in all investigated samples and low-field non-resonance signal of microwave absorption in the superconducting phase are detected. The effective values of g-tensor and minimal EPR linewidths of $Yb^{3+}$ ions in parallel (H∥***C***) orientation are presented in Table 1. The experimental g-values are in an excellent agreement with calculations using the crystalline electric field parameters as determined by inelastic neutron scattering [4, 5]. The integrated intensity of EPR signal in the superconducting species is increasing with decreasing temperature below 80 K but slower than it would be expected according to the Curie law.

Table I. Critical and Debye temperatures, the minimal EPR linewidth $\Delta H_{PP}^{min}$ measured at temperatures $T_{min}$, effective g-values[a], and the SLR parameter $C$ against oxygen content X in $YBa_2Cu_3O_x$ compounds.

| X | 6.85 | 6.67 | 6.45 | 6.0 |
|---|------|------|------|-----|
| $T_C$ (K) | 85 | 65 | 40 | - |
| $T_{min}$ (K) | 55 K | 45 K | 25 K | 20 K |
| $\Delta H_{PP}^{min}$ (G) | 95 | 90 | 75 | 95 |
| $g$ | 3.07 | 3.11 | 3.18 | 3.48[b] |
| $\Theta_D$ (K) | 450 | 370 | 280 | 250 |
| $C$ ($10^{-7} s^{-1} K^{-9}$) | 1.06 | 4.23 | 15.3 | 50 |

[a] The g-values were estimated at T = 50 K.

[b] For the sample with X = 6.0 the perpendicular value of g-tensor $g_\perp$ is given

Fig. 2 shows the complicated temperature dependence of the $Yb^{3+}$ EPR peak-to-peak linewidth $\Delta H_{PP}$ in $YBa_2Cu_3O_{6.85}$ with a minimum at the temperature $T_{min}$ and a small step of $\Delta H_{PP}$ near $T_C$. The EPR measurements on this sample at $T < T_{min}$ were discussed previously [6] and not in the scope of the present paper. A rapid increase of $\Delta H_{PP}$ as T increases (non-linear, non-Korringa slope) in all investigated samples for $T > T_{min}$ is very common for the rare-earth ions in single crystals (except for these with half-filled *f*-shell e.g. $Gd^{3+}$) and, as we have shown for $Yb^{3+}$ and $Er^{3+}$ probes in YBaCuO compounds [6, 7, 8], caused by the *4f*-electron–phonon interaction only. To determine the dominating spin-lattice relaxation process we use the next approach.

The EPR linewidth caused only by the SLR, $\Delta H_{PP}^{SLR}$, can be extracted from the experimentally measured EPR linewidth $\Delta H_{PP}$ by the relation [9]

$$\left(\Delta H_{PP}\right)^2 = \Delta H_{PP}^{SLR} \cdot \Delta H_{PP} + \left(\Delta H_{PP}^{min}\right)^2. \qquad (1)$$

(We take into account the fact that the EPR line has a Gaussian lineshape at $T = T_{min}$ and arises due to inhomogeneities of the crystal electric field (CEF) potential, and the Lorentzian lineshape is connected usually to the term $\Delta H_{PP}^{SLR}$. The extracted data just slightly differ from those obtained by using the well-known expressions $\Delta H_{PP} = \Delta H_{PP}^{SLR} + \Delta H_{PP}^{min}$ for Lorentzian-Lorentzian or $(\Delta H_{PP})^2 = (\Delta H_{PP}^{SLR})^2 + (\Delta H_{PP}^{min})^2$ for Gaussian-Gaussian convolutions, correspondingly [10]).

The SLR rate $T_1^{-1}$ can be expressed through $\Delta H_{PP}^{SLR}$ as follows [11]:

$$T_1^{-1}(s^{-1}) = (\sqrt{3}/2) g \beta \hbar^{-1} \cdot \Delta H_{PP}^{SLR}(G) \equiv 7.62 \cdot 10^6 g \cdot \Delta H_{PP}^{SLR}(G), \qquad (2)$$

where $\beta$ is Bohr's magneton, and $\hbar$, the Planck constant.

Fig. 3 shows the temperature dependences of the SLR rate $T_1^{-1}$ for the samples with different oxygen contents. The relaxation rate $T_1^{-1}$ of $Yb^{3+}$ ions can be described approximately using formula [12]

$$T_{1R}^{-1} = C \cdot T^9 \cdot f(\Theta_D/T), \qquad (3)$$

where $f(\Theta_D/T) \equiv f(z) \equiv I_8(z)/I_8(\infty)$, $I_8(z) = \int_0^z x^8 e^x/(e^x - 1)^2 dx$, $I_8(\infty) = 8!$

(dashed line in Fig. 4). This corresponds to the usual Raman SLR process with Debye approximation for lattice vibrations (and, therefore, label $T_{1R}^{-1}$ for the SLR rate in Eq. (3) is used). The magnitudes of the Debye characteristic temperature $\Theta_D$ and factor $C$ are listed in Table 1. The values of $\Theta_D$ are in sufficiently good agreement with the data obtained from the specific heat capacity measurements [13] and calculations based on measurements of the elastic constants [14, 15].

The more detailed analysis is given in our previous paper [7] but we did not discuss the SLR rate anomaly near $T_C$ there.

The temperature dependence of $T_1^{-1}$ can be alternatively described as [16]

$$T_1^{-1} = B \cdot \exp(-\Delta_{SLR}/T) \qquad (4)$$

(solid lines in Fig. 4). It manifests clearly the $T_C$ anomaly observed. For the sample with X = 6.85, for example, the factor $B$ decreases approximately twice with temperature ($B_1 \approx 11.4 \cdot 10^{11}$ at T < 80 K and $B_2 \approx 5.9 \cdot 10^{11}$ at T > 100 K with $\Delta_{SLR} = 500$ K). The lower the oxygen content is the weaker the SLR rate anomaly reveals.

The exponential description in Eq. (4) may be assigned to the two-stage relaxation process via an intermediate energy level as well as to the Raman process with participation of some separated (optical, local, etc) vibrations in the crystal. The values of $\Delta_{SLR}$ are very close to those of $\Theta_D$ for all investigated samples and do not correlate with the energy of the first excited level of $Yb^{3+}$ ions in YBCO compounds ($\approx$ 1000 K [17], practically independent of oxygen content). Furthemore, for the two-step relaxation process with the Stark exitation energy of about 500 K, the order of magnitude of pre-exponential factor $B$ should be approximately $10^{15}$. It is much more than $\approx 10^{12}$ estimated from our data. Therefore, we can exclude the two-stage relaxation process and conclude anew that the line broadening of $Yb^{3+}$ ions with temperature is caused by the Raman SLR process.

We have to note that the values of $\Delta_{SLR}$ (or $\Theta_D$) are in good correspondence with the peaks positions in Raman and inelastic neutron scattering spectra of high-$T_C$ YBCO compounds. So authors of paper [18] discuss the origin of double peak observed near 300 $cm^{-1}$ ($\approx$ 430 K) in $NdBa_2Cu_3O_7$. Pronounced enhancements of the spectral weight centered around 40 meV ($\approx$ 460 K) for X = 7.0, 33 meV ($\approx$ 380 K) for X = 6.7, and 25 meV ($\approx$ 290 K) for X = 6.5 below $T_C$ in $YBa_2Cu_3O_X$ are revealed in [19] (cf. with the values of $\Theta_D$ in Table 1).

The anomaly of the spin-lattice relaxation rate near $T_C$ can be associated with the differences of the phonon spectrum of the YBaCuO host matrix in the superconducting and non-superconducting states. For example, as it is shown in [20], phonon spectra density of the superconducting sample with X = 6.95 at T = 80 K drastically differs from that at T = 290 K, while there is no great difference for the sample with X = 6.35. It is sensible for $Yb^{3+}$ ions because practically whole phonon spectrum is involved into the usual Raman relaxation process. In addition, we have to note again that a local peak in the vicinity of 40 meV ($\approx$ 460 K) also observed in [20] for the superconducting species.

We did not obtain the $T_C$ anomaly on the $Er^{3+}$ doped ions in $Er_{0.01}Y_{0.99}Ba_2Cu_3O_{6+X}$ compounds in our experiments [8]. It can be also understood in the frames of the above approach. The line broadening of $Er^{3+}$ ions with temperature is caused by the two-step resonance fluorescence SLR process. The phonon density for the phonons with energy of 9-12 meV (corresponding to Stark exitation energy $\Delta$ for $Er^{3+}$) involved into the relaxation mechanism is approximately constant below and above $T_C$ [20].

Our attempts to associate the anomaly with the small changes of the sound velocity $\upsilon$ in the superconducting species (see [21], for example) do not give any reasonable quantitative or qualitative result though the value $\upsilon$ appears as $\upsilon^{-10}$ in the coefficient $C$ for usual Raman and as $\upsilon^{-5}$ in the coefficient $B$ for two-step processes, correspondingly. Although the presumption could be right, it would be quite difficult to explain the absence of the $T_C$ anomaly for $Er^{3+}$ ions.

## Conclusion

The spin-lattice relaxation rate measurements of doped $Yb^{3+}$ ions from the temperature dependence of electron spin resonance linewidth in $Y_{0.99}Yb_{0.01}Ba_2Cu_3O_X$ with different oxygen contents have been made. Near $T_C$ the small step in the EPR linewidth and corresponding anomaly in the $T_1^{-1}$ have been observed. The anomaly can be ascribed to the phonon spectrum density changes of YBaCuO host matrix in the superconducting and non-superconducting states. In this connection it might be interesting to carry out such experiments with other rare-earth impurities and in other frequency ranges.

## Acknowledgments

This work was supported by CDRF grant REC-007, scientific programm "Universities of Russia", and scientific Fund of Republic of Tatarstan (NIOKR RT), Russia.

## Figure captions

Fig. 1. The EPR spectrum of $Yb^{3+}$ ions in $Y_{0.99}Yb_{0.01}Ba_2Cu_3O_{6.67}$

Fig. 2. Temperature dependence of the peak-to-peak EPR linewidth of $Yb^{3+}$ ions in $YBa_2Cu_3O_{6.85}$, $H \| C$

Fig. 3. SLR rates for $Yb^{3+}$ ions in $YBa_2Cu_3O_X$ with $X = 6.85$ (a), 6.67 (b), and 6.0 (c) derived from the EPR line broadening using Eq. (2). The data are presented in log-log scale.

Fig. 4. Temperature dependence of SLR rate of $Yb^{3+}$ ions in $YBa_2Cu_3O_{6.85}$. Dashed line – approximation using Eq. (3) with parameters listed in Table 1. Solid lines – exponential approximation using Eq. (4) with $\Delta = 500$ K, $B_1 = 11.4 \cdot 10^{11}$ s$^{-1}$ (below $T_C$), and $B_2 = 5.9 \cdot 10^{11}$ s$^{-1}$ (above $T_C$). The data are presented in logarithmic scale.

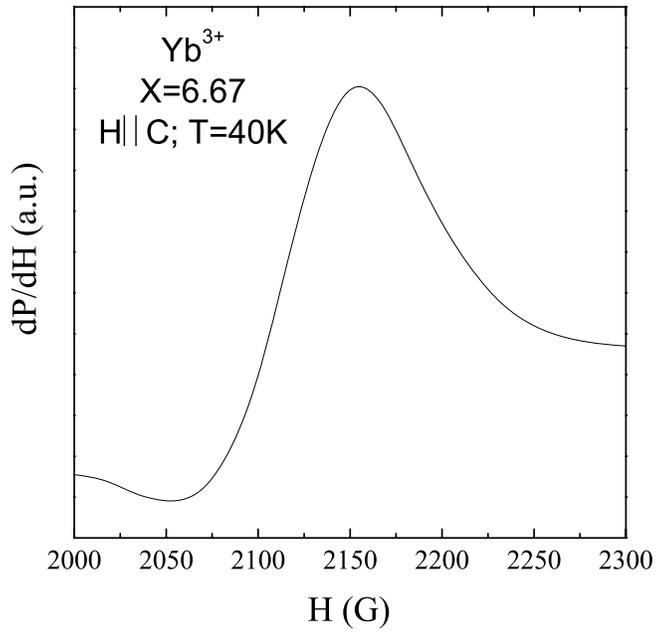

Fig.1

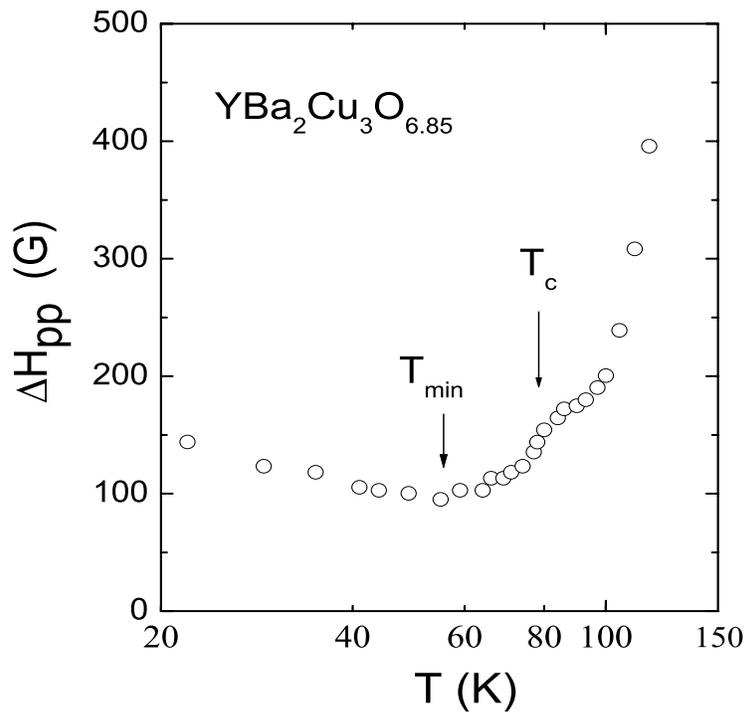

Fig. 2.

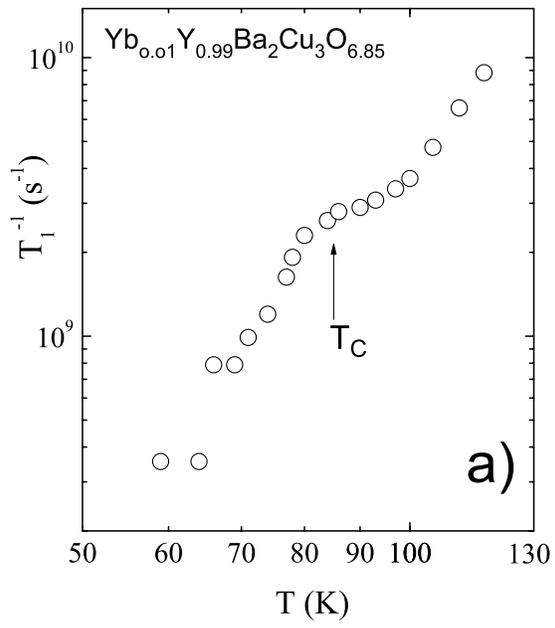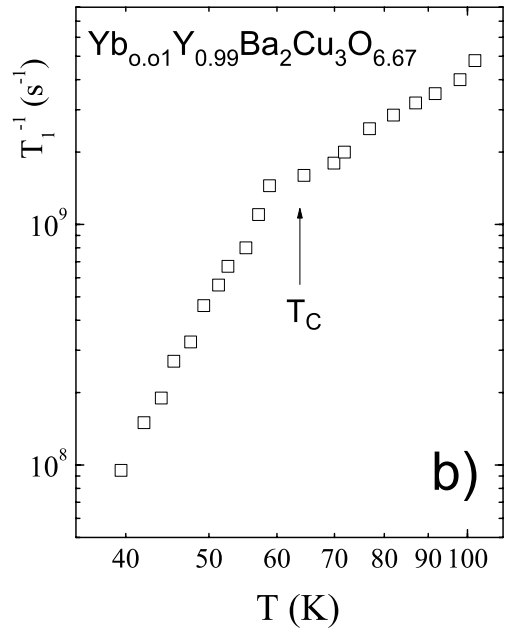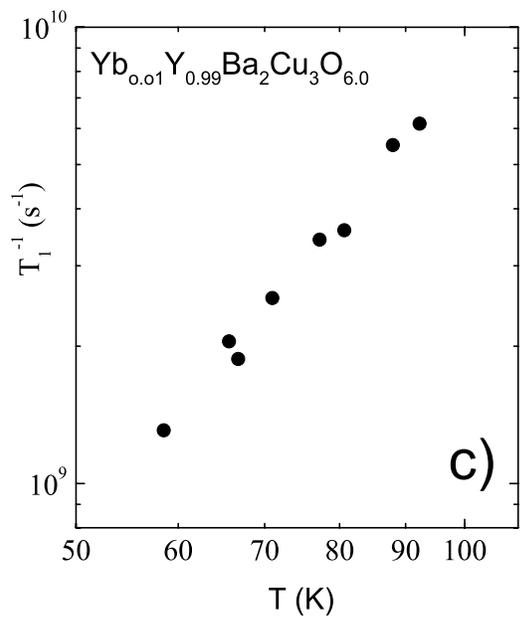

Fig.3.

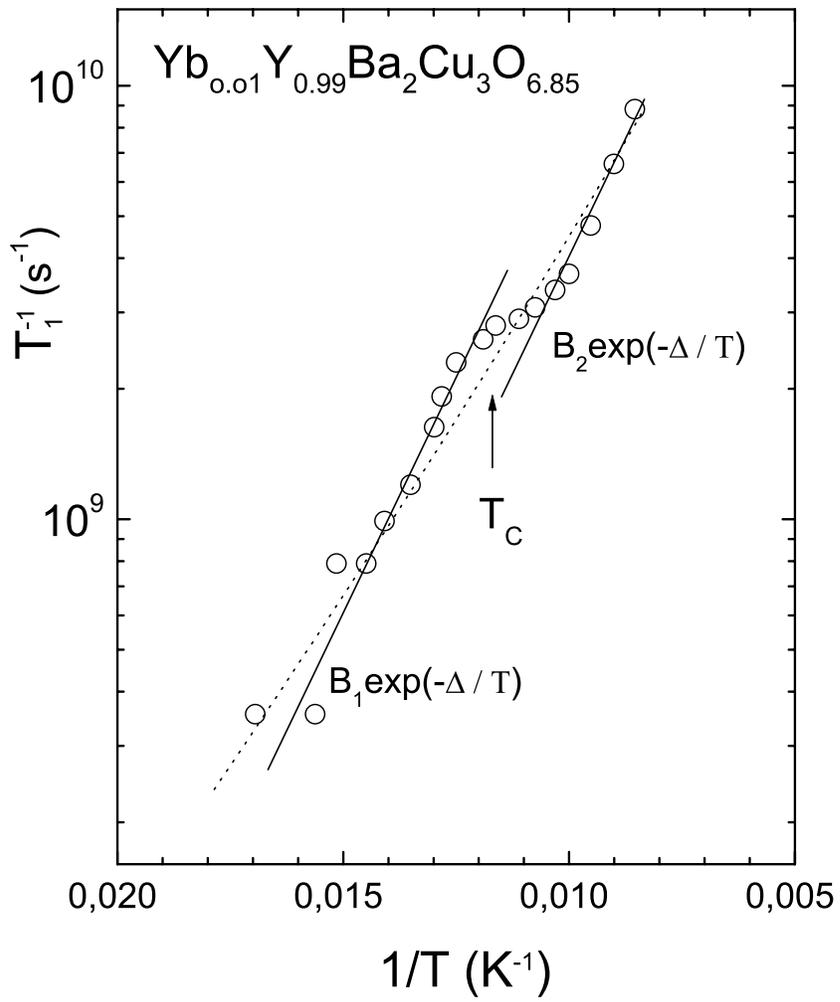

Fig.4.